\documentclass[dvips]{article}

\usepackage{icrc2011}

\title{MAGIC observations of the giant radio galaxy M87 in a low emission state between 2005 and 2007}

\newcommand{\etal}{\MakeLowercase{\textit{et al. }}} 
\shorttitle{Berger \etal M87}

\authors{Karsten Berger$^{1,2}$, Dijana Dominis Prester$^{3}$, Fabrizio Tavecchio
$^{4}$, Tomislav Terzi\'c$^{3}$, MAGIC Collaboration$^{5}$ }
\afiliations{$^1$Inst. de Astrof\'{\i}sica de Canarias, E-38200 La Laguna, Tenerife, Spain\\ $^2$Depto. de Astrof\'{\i}sica, Universidad de La Laguna, E-38206 La Laguna, Spain\\$^3$University of Rijeka, HR-51000 Rijeka, Croatia\\ $^4$INAF National Institute for Astrophysics, I-00136 Rome, Italy\\ $^5$The full list of collaborators can be found at: wwwmagic.mppmu.mpg.de }
\email{berger@astro.uni-wuerzburg.de}

\abstract{We present the results of a long M87 monitoring campaign in very high energy $\gamma$-rays with the MAGIC-I Cherenkov telescope. A total of 150 hours of data was gathered between 2005 and 2007. No flaring activity was found during that time. Never the less, we have found an apparently steady and weak signal at the level of $7\sigma$. We present the spectrum between 100 GeV and 2 TeV, which is consistent with a simple power law with a spectral index $-2.21\pm0.21$ and a flux normalization (at 300 GeV) of $5.4\pm1.1 \times 10^{-8} \frac{1}{\mathrm{TeV s m}^{2}}$. It complements well with the previously published Fermi spectrum, covering an energy range of four orders of magnitude without apparent change in the spectral index. }

\begin{document}
\maketitle

\section{Introduction}


M87 is a giant elliptical radio galaxy situated in the Virgo cluster at a distance of 16.7 Mpc \cite{lab41}. The centre of the galaxy is a SMBH with a mass of $(6.4 \pm 0.5)\times 10^9 M_\odot$ \cite{lab26}.
The prominent jet of M87 (first discovered in the optical band \cite{lab24}) is inclined by $10^\circ - 45^\circ$ from our line of sight \cite{lab18, lab37}. The jet is not homogeneous but shows several distinctive regions.
The compact nucleus has a dimension of several hundreds of Schwarzschild radii. Throughout the inner jet several brighter spots, so called ``knots", are visible, often characterized by superluminal motion with apparent speeds reaching $\sim6c$ \cite{lab18}. The jet continues to expand into the inter-cluster medium, forming giant lobes with a length of 80 kpc. A closer inspection also indicates the existence of a counter-jet.
The large scale jet primarily emits from radio to X-rays through the synchrotron mechanism. The two most promising emission regions at all wavelengths (including very high energy $\gamma$-rays) are the nucleus and the knot closest to the nucleus (60 pc), known as HST--1.


The first indication that M87 is a source of VHE $\gamma$-radiation was reported by the HEGRA collaboration \cite{lab5}, which was confirmed by the H.E.S.S collaboration with data collected between 2003 and 2006 \cite{lab6}.
This contribution will focus on long-term monitoring observations of M87 with MAGIC during a low emission state in 2005--2007.

\section{Observations and Data Analysis}

MAGIC started to observe M87 regularly in 2005 and continued to monitor it during several observational cycles until today (see e.g. the recent announcement of an enhanced emission state in February 2010 \cite{lab38}). This paper focuses on data taken between March 2005 and June 2007 with the stand-alone MAGIC-I telescope. The MAGIC-I telescope underwent several significant upgrades over this period. Specifically these upgrades include:
The optical PSF of the 17m mirror dish improvement.
The readout chain has been significantly enhanced during the observation time. The data discussed in this paper include three different readout stages: 1) 300\,MS/s readout without splitters, 2) 300\,MS/s readout including splitters and 3) 2\,GS/s readout with splitters.
After the installation of the 2\,GS/s Mux-FADC system a detailed study of the timing properties of the showers lead to a significant improvement of the sensitivity of the telescope \cite{lab14}.

The analysis technique used throughout this paper is the same one performed in \cite{lab11}, where an excellent agreement between the published Crab Nebula spectrum and the stacked analysis was found, confirming the stability of the analysis over large time scales. 

MAGIC observations of M87 were performed in two different modes: ON-OFF and wobble. In the ON-OFF mode the telescope is pointed directly at the source during the ON runs. The background is estimated with dedicated OFF data.
In the so-called wobble mode \cite{lab25} the telescope is not pointed directly at the source position but 0.4$^{\circ}$ away.  Accordingly wobble data does not require dedicated OFF runs.
Only the 2005 observations of M87 were recorded in ON-OFF mode while the majority of the data taken in the later years was taken in wobble mode.
Throughout this paper we have applied the same timing based analysis to the entire dataset, which is described in detail in \cite{lab10}. Each period with different telescope conditions was first analysed separately with matching OFF-data (when needed) and Monte Carlo simulations from the corresponding period and finally combined after the $\gamma$--hadron separation.

The entire observation campaign consists of 154.1\,h out of which 128.6\,h survive the quality selection cuts. This selection ensures that only data with good weather conditions and without any technical problems is used in the analysis. Table \ref{tab:obs.periods} gives an overview how this data is distributed over the entire campaign.

\begin{table}
\begin{center}
\begin{tabular}{| c | c | c | c | c |}
  \hline
  Obs & Obs & read-out & PSF & Obs  \\
  period & mode & system & & time \\ \hline
  2005a & On-Off & 300\,MS/s & 20\,mm & 6.5\,h \\ \hline
  2005b & On-Off & 300\,MS/s & 13\,mm & 1.5\,h \\ \hline
  2006a & Wobble & 300\,MS/s & 15\,mm & 53.4\,h \\ \hline
  2006b & Wobble & 300\,MS/s + & 14\,mm & 26.0\,h \\
  & & Splitters & & \\ \hline
  2007 & Wobble & 2\,GS/s + & 14\,mm & 42.8\,h \\
  & & Splitters & & \\ \hline
 \hline
\end{tabular}
\caption{Observation periods}
\label{tab:obs.periods}
\end{center}
\end{table}

\section{MAGIC results}

\begin{figure}[!t]
\includegraphics[angle=270,width=0.45\textwidth]{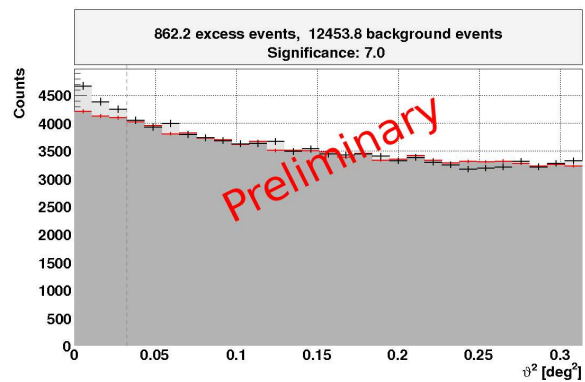}
\caption{$\theta^2$ distribution of the combined 128.6\,h MAGIC observations between 2005 and 2007. The grey shaded area below the red crosses is the OFF data-sample, while the black crosses corresponds to the ON data. The excess is point-like compared to the PSF with a significance of 7 standard deviations (calculated with formula 17 of \cite{lab35}). }
\label{fig:theta2}
\end{figure}

The $\theta^2$ distribution of the entire observation campaign is shown in Fig. \ref{fig:theta2}. An excess of 862 events over 12454 background events is apparent, corresponding to a significance of 7.0 $\sigma$ (calculated with formula 17 of \cite{lab35}).

\begin{figure}[!t]
\includegraphics[angle=270,width=0.45\textwidth]{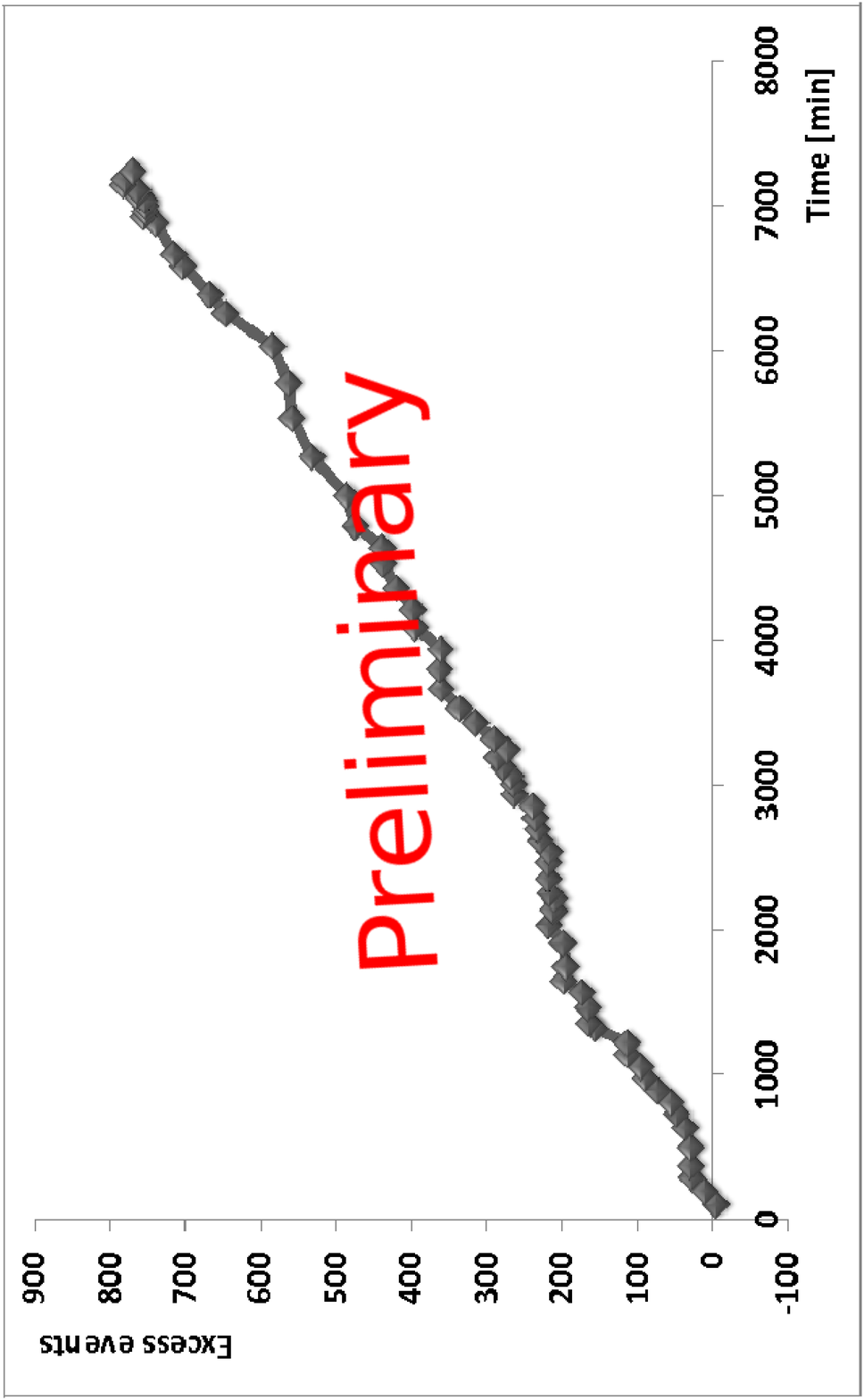}
\caption{Evolution of excess events from M87 over time. Only wobble data are shown in this figure in order to reduce systematic uncertainties in the ON-OFF subtraction.}
\label{fig:excess}
\end{figure}

\begin{figure}
\includegraphics[width=0.45\textwidth]{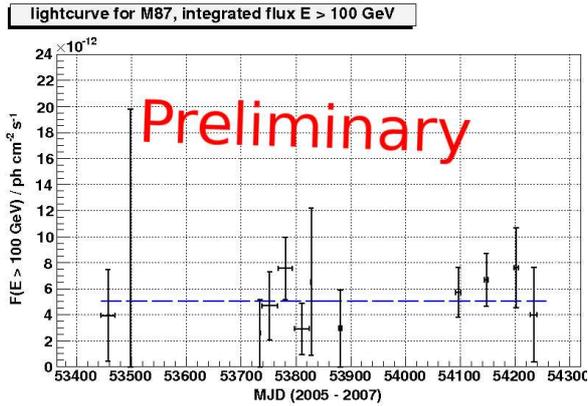}
\caption{Light curve of the integral $\gamma$-ray flux above 100\,GeV during the 2005--2007 MAGIC observation period. The dashed blue line corresponds to the fit result of a linear function to the data points, with a reduced $\chi^2/d.o.f. = 5.66/11 = 0.51$ and a mean flux of $5.06\times 10^{-12} \frac{1}{\mathrm{TeV s cm}^{2}}$.}
\label{fig:LC}
\end{figure}

As the light curve in Fig. \ref{fig:LC} shows, the excess over the entire observation campaign is compatible with a fit to a constant flux with a reduced $\chi^2$ of 0.51. Additionally we analysed each observation night independently, however none of the nights yielded an excess $>$3$\sigma$, which is consistent with random fluctuations. Fig. \ref{fig:excess} shows the cumulative excess distribution over time. The excess grows linearly, which is consistent with a constant emission. We thus conclude that no significant flare occurred during our observation campaign. Note that the data in 2005 are compatible within errors with the high flux state observed by H.E.S.S. \cite{lab6}. The reason is mainly a too short MAGIC observation time (and accordingly large error on the source flux) as well as the short time variability of the source (flares on a time scale of two days have been reported by the H.E.S.S. Collaboration).

\begin{figure}[!t]
\includegraphics[width=0.45\textwidth]{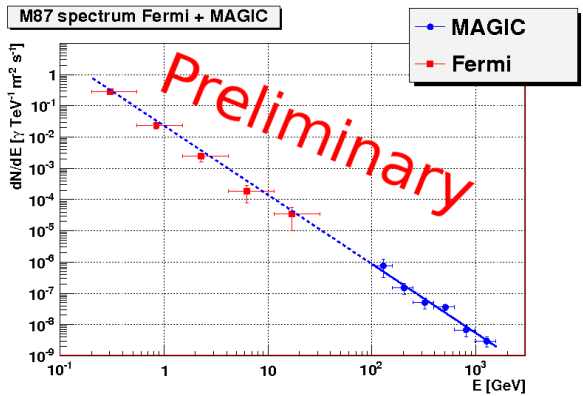}
\caption{Shown above is the combined \textit{Fermi}--MAGIC differential energy spectrum over four orders of magnitude in energy starting from 100\,MeV up until 1.5\,TeV. The dashed line corresponds to the extrapolation of the fit to the MAGIC data points into the \textit{Fermi} energy range. Note that the two observations are not contemporaneous.}
\label{fig:spec}
\end{figure}

The MAGIC spectrum is consistent with a simple power law with a spectral index -2.21$\pm$0.21 and a flux normalization (at 300 GeV) of $5.4\pm1.1 \times 10^{-8} \frac{1}{\mathrm{TeV s m}^{2}}$. Both the \textit{Fermi} spectrum (taken from \cite{lab1}) and the MAGIC spectrum are shown in Fig. \ref{fig:spec}, covering an energy range of four orders of magnitude without apparent change in the spectral index.

A detailed discussion of the potential systematic errors can be found in \cite{lab8}. As a summary the systematic energy scale error is about 16$\%$, the error of the flux normalization is estimated to be 11$\%$ and the systematic slope error is $\pm$0.2. Using Monte Carlo simulations for each observational epoch we find that the error of the flux normalization should be $<$10$\%$ and the error of the slope should be $<$0.03 (note that this applies only to hard spectra with a spectral index $<$2.5 as is the case for the data discussed here).

\section{Modelling the SED}

We model the SED using the structured jet model of \cite{lab28}, already applied to M87 in \cite{lab47}. The model assumes that the jet has an inner fast core (spine), with bulk Lorentz factor $\Gamma_{\rm S}$, surrounded by a slower layer, with bulk Lorentz factor $\Gamma_{\rm L}$. In both regions, relativistic electrons emit through synchrotron and inverse Compton mechanisms.

The model is fully specified by the following parameters: {\it i)} the spine is assumed to be a cylinder of radius $R$, height $H_{\rm S}$ (as measured in the spine frame) and in motion with bulk Lorentz factor $\Gamma_{\rm S}$; {\it ii)} the layer is modelled as a hollow cylinder with internal radius $R$, external radius $R_2$, height $H_{\rm L}$ (as measured in the frame of the layer) and bulk Lorentz factor $\Gamma_{\rm L}$. Each region contains tangled magnetic field with intensity $B_{\rm S}$, $B_{\rm L}$ and it is filled by relativistic electrons assumed to follow a (purely phenomenological) smoothed broken power--law distribution extending from $\gamma_{\rm min}$ to $\gamma_{\rm max}$ and with indices $n_1$, $n_2$ below and above the break at $\gamma_{\rm b}$.  The normalisation of this distribution is calculated assuming that the system produces an assumed (bolometric) synchrotron luminosity $L_{\rm syn}$ (as measured in the local frame), which is an input parameter of the model. We assume that $H_{\rm L} > H_{\rm S}$. The seed photons for the IC scattering are not only those produced locally in the spine (layer), but we also consider the photons produced in the layer (spine).

\begin{figure}
\includegraphics[width=0.45\textwidth]{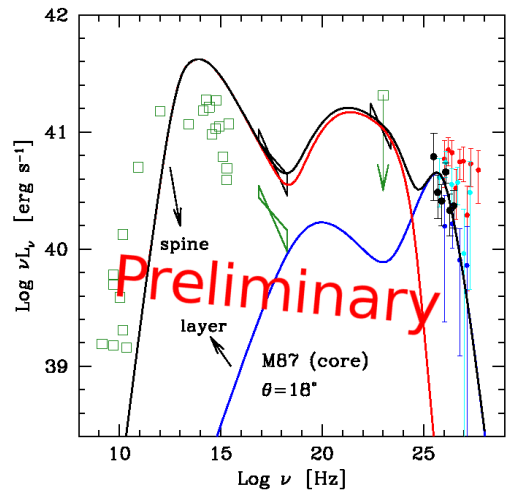}
\caption{SED of the core of M87 (green open squares) together with the MAGIC data points (black filled circles). Refer to the text for details.}
\label{fig:mod}
\end{figure}

The result of the modelling is reported in Fig. \ref{fig:mod}, where we show the SED of the emission produced by the spine (red) and the layer (blue) and their sum (black). Green open symbols report historical data, black filled circles show the MAGIC spectrum. For comparison we also report the H.E.S.S. spectra taken in 2004 and 2005 (blue and red circles, respectively, from \cite{lab6}) and the 2007 VERITAS spectrum (cyan, from \cite{lab2}). Clearly the 2005--2007 spectrum was very similar to that measured by H.E.S.S. in the ``low" 2004 state and by VERITAS. The green ``bow-tie" reports the nuclear X-ray spectrum measured by \textit{Chandra} in 2000 \cite{lab16}. The black ``bow-tie" show the average X-ray spectrum during the period covered by the MAGIC observations, obtained by scaling the 2000 spectrum by a factor of 4, as inferred in from the lightcurve reported by \cite{lab32}. During the MAGIC observing periods there are no measurements in the GeV energy band. However, for comparison, we report the \textit{Fermi}/LAT spectrum obtained integrating over the first 10 months of all-sky survey data \cite{lab1} with a black bow-tie. The green open square in the high energy $\gamma$-ray range corresponds to the EGRET upper limit. The SED shows two pronounced bumps, one peaking in the IR band, one extending from MeV to TeV energies. The adopted parameters are reported in Table \ref{tab:modelparam}.

\begin{table*}
\begin{center}
\begin{tabular}{|l|llllllllllll|}
\hline
& $R$    
& $H$  
& $L_{\rm syn}$  
& $B$  
& $\gamma_{\rm min} $  
& $\gamma_{\rm b} $ 
& $\gamma_{\rm cut}$  
& $n_1$
& $n_2$
& $\Gamma$ 
& $\theta$ & \\
&cm  &cm &erg s$^{-1}$  &G & & & & & & &deg. & \\
\hline  
spine  &7.5e15 &3e15 &4.7e41 &2.1    &600 &2e3 &1e8 &2 &3.65 &12  &18 &\\ 
layer  &7.5e15 &6e16 &1.6e38 &0.35 &1   &2e6 &1e9 &2 &3.3 &4  &18  &\\
\hline
\end{tabular}
\caption{Input parameters of the models for the layer and the spine shown in Fig. \ref{fig:mod}. All quantities (except the bulk Lorentz factors $\Gamma$ and the viewing angle $\theta$) are measured in the rest frame of the emitting plasma. The external radius of the layer is fixed to the value $R_2=1.2 \times R$.}
\label{tab:modelparam}
\end{center}
\end{table*}       

A feature of the realization of the ``spine-layer" model for M87 is that the IC emission from the layer, accounting for the observed VHE emission, is partly absorbed through the interaction with the optical-IR photons produced in the spine. This leads to a relatively soft spectrum, suitable to reproduce the low level spectrum measured by MAGIC, but difficult to reconcile with the hard spectrum recorded during high state by H.E.S.S. A possible way out of this problem is to enlarge the emission regions. However, the increase of the source size is limited by the observed short variability time-scales.

It appears that the power associated to the magnetic field is largely dominant over that related to relativistic electrons and protons (assuming a composition of 1 cold proton per emitting electron, see e.g. \cite{lab29}) both in the jet and in the layer. In any case the jet power required to reproduce the SED, strongly dominated by the spine, appears relatively modest, $P_{\rm jet}=1.5\times 10^{44}$ erg/s.

\section{Summary}

MAGIC has detected weak and steady VHE $\gamma$-ray emission from M87 between 2005 and 2007. Our measurements are compatible with previously reported low states by VERITAS and H.E.S.S., which suggests that the observed emission level and spectral characteristics are extremely stable and may refer to the "ground-state" emission of M87. This is also supported by the extrapolation of the MAGIC spectrum into the \textit{Fermi} energy range, which is in perfect agreement despite the fact that both observations are not contemporaneous.

We were able to describe this emission with a structured jet model, that separates the jet into a spine and an outer layer.
An interesting feature of the model is the existence of a transition region between the emission from the spine and that from the layer around 40\,GeV. Currently no measurements of M87 are available in this energy range. This is due to the small collection area of the \textit{Fermi} satellite and the insufficient sensitivity of Cherenkov telescopes in the low energy domain. Since fall 2009 MAGIC consists of two 17\,m diameter telescopes, which observe in a stereoscopic mode. This upgrade has especially improved the instrument's capabilities at energies below 100\,GeV \cite{lab13}. It is thus possible that a future, deep observation of M87 with the stereoscopic MAGIC system will reveal a feature in the otherwise smooth power law spectrum of M87, confirming or denying the validity of the structured jet model.


\clearpage

\end{document}